\documentclass[twocolumn,showpacs,preprintnumbers,nofootinbib,amsmath,amssymb]{revtex4}
\begin{document}
%\draft

\title{Multiple M$0$-brane equations in eleven dimensional pp-wave superspace and  BMN matrix model}

%Berenstein--Maldacena--Nastase

\author{Igor A. Bandos $^{\dagger\ddagger}$
}
\address{$^{\dagger}$Department of
Theoretical Physics, University of the Basque Country UPV/EHU,
% Barrio Sarriena s/n, 48940 Leioa, Spain
% (Vizcaya)
P.O. Box 644, 48080 Bilbao, Spain
 \\ $^{\ddagger}$
IKERBASQUE, Basque Foundation for Science, 48011, Bilbao, Spain}

\date{February 24, 2012}

\def\theequation{\arabic{section}.\arabic{equation}}

\begin{abstract}

We obtain the Matrix model equations in the  background of the maximally supersymmetric pp-wave  solution of the 11D supergravity and discuss its relation with the Berenstein--Maldacena--Nastase (BMN)  model.

\end{abstract}


\maketitle

\section{Introduction}

Matrix model \cite{Banks:1996vh} was proposed 15 years ago and all these years
was an important tool for studying M-theory \cite{M-theory}, see e.g. \cite{Susskind:1997cw,Seiberg:1997ad,Emparan:1997rt,BMN,pp-w=DSR,SheikhJabbari:2004ik,Sadri:2003pr,Verlinde+C+S=2005,Craps:2006yb,Nastase-ABJM}. It was conjectured to provide a non-perturbative description of M-theory in some limit and,
to stress this, the name M(atrix) theory was often used. Although this M(atrix) theory was considered to be eleven dimensional, the Lagrangian staying beyond it according to \cite{Banks:1996vh} was just a dimensional reduction of D=10 SYM down to d=1 (which for the gauge group $U(N)$ is believed to provide a low energy description of the  system of $N$ nearly--coincident D$0$-branes);  the  symmetry enlargement to include D=11 Lorentz group was discussed in \cite{Banks:1996vh} and later papers. However it was not clear how to write the action for Matrix model in 11D supergravity background. This is why Matrix model action (and equations of motion) are known for a few particular supergravity backgrounds only, and for these they were rather guessed than derived. In particular, the action for Matrix model in maximally supersymmetric pp-wave  background was proposed by Berenstein, Maldacena and Nastase in 2002 \cite{BMN} and is known under the name of BMN (matrix) model. The other example is the so--called Matrix Big Bang background \cite{Verlinde+C+S=2005,Craps:2006yb}.

A natural way to resolve the problem was to obtain invariant action, or covariant equations of motion, for multiple M0-brane system. Indeed, the Matrix model originally was conjectured to be the theory of nearly coincident multiple D$0$--branes (mD$0$) \cite{Banks:1996vh} so that, as far as single D$0$-brane can be obtained by dimensional reduction of M$0$--brane \cite{B+T=Dpac}, it is natural to expect that multiple M$0$--brane  system (mM$0$) stays beyond the conjectured enlargement of the 10D Lorentz group symmetry of mD$0$ till 11D Lorentz group $SO(1,10)$.

However, to write Lorentz invariant and supersymmetric  action for mM$0$ system was not so easy: it was a particular case of the problem which in D=10 type II case was  known as a search for supersymmetric non-Abelian Born-Infeld action \cite{Tseytlin:DBInA}. The progress towards solution of this problem is only partial, although impressive. We refer on our previous papers \cite{mM0=PLB,mM0=PRL,mM0=PRD} for the elaborated description of the results of refs. \cite{Myers:1999ps,Dima01,Howe+Linstrom+Linus,YLozano+=0207,BLG,ABJM,IB09:D0}
(the list of which is certainly incomplete), but here restrict our discussion by stating what is known about multiple M$0$-brane (mM$0$) system. The purely bosonic action for
mM$0$ was proposed in \cite{YLozano+=0207}. However, as far as this was the straightforward counterpart of the 'dielectric brane action' proposed by Myers for multiple D$p$-branes \cite{Myers:1999ps}, it did not possess neither supersymmetry nor 11D Lorentz symmetry (nor complete diffeomorphism symmetry), so that it was not clear how to introduce the coupling to the 11D supergravity background in such an action.

To resolve the problem of covariant and supersymmetric description of multiple mM$0$ system, the superembedding  approach to this system was proposed in  \cite{mM0=PLB}\footnote{Superembedding approach for superstrings and eleven dimensional supermembrane was proposed in \cite{bpstv} developing the line of the so-called STV (Sorokin-Tkach-Volkov) approach to superparticle \cite{stv} and superstring \cite{DGHS93} (see \cite{Dima99} for more references on STV approach), and was generalized to Dirichlet branes (D$p$-branes) in \cite{hs96} (see also \cite{bst97}) and to M5-brane in \cite{hs2}, respectively. See \cite{Dima99,IB09:M-D} for reviews and more references.}. It was shown in \cite{mM0=PLB} that, for the case of flat target superspace, the equations of relative motion of the multiple M$0$-brane constituents, which follows from the proposed superembedding approach equations, coincide with the Matrix model equations of \cite{Banks:1996vh}. The superembedding approach to the mM$0$ system in an arbitrary curved 11D supergravity superspace was developed in \cite{mM0=PRL,mM0=PRD}, where the equations of motion, which can be treated as equations of Matrix model coupled to an arbitrary supergravity  background, were obtained.

The natural application of the results of \cite{mM0=PRL,mM0=PRD} is to obtain the Matrix model equations in particular interesting M-theoretical background, like $AdS_4\times S^7$ and $AdS_7\times S^4$ and to apply them in  context of the AdS/CFT correspondence. It happens however that  this is not so easy even for bosonic backgrounds: as we will see below on a relatively simple example, the final result cannot be reached by just substituting a particular bosonic solution of the spacetime 11D supergravity equations in the general Matrix model equations of \cite{mM0=PRL,mM0=PRD}, but requires first to lift of that bosonic solutions to a curved superspace solving the superspace supergravity constraints.  Furthermore, then we have to find certain information about the geometry of the worldline superspace, describing the center of energy motion of the mM$0$ system, embedded into that particular supergravity superspace.

Thus it is natural to begin the program of specifying the Matrix model equations of \cite{mM0=PRL,mM0=PRD} for particular 11D supergravity backgrounds with studying the case of maximally supersymmetric pp--wave background which is simpler than $AdS\times S$ ones, but appears as a Penrose limit of these backgrounds. Furthermore, the BMN matrix model was proposed as a candidate for the M(atrix) model in that background, so that, on this way, we can check whether the BMN conjecture \cite{BMN} was correct. (Alternatively, putting the previous statement  bottom up to make it acceptable for strong believers of the BMN model, such a study could be considered as checking the equations obtained in \cite{mM0=PRL,mM0=PRD}, using the BMN model as a reference point).

The derivation of the matrix model equations in supersymmetric pp-wave background,
starting from the results of \cite{mM0=PRL,mM0=PRD}, and their comparison with the BMN
equation is the subject of this paper.
We find a vacuum solution of the equations of the center of energy motion of the mM$0$ system, in which all the Goldstone fields corresponding to the symmetries and supersymmetries broken spontaneously by the center of energy motion of mM$0$ system are equal to zero. We describe this solution as a worldline superspace and use this to obtain  the equations of the  relative motion of the constituents of mM$0$ system moving in the 11D pp-wave superspace. These equations indeed coincide with the ones of the SU(N) sector of the BMN model. Furthermore, we show that the complete set of mM$0$ equations coincide with the equations of the  BMN model in the leading order on the Goldstone fields of the center of energy motion.

The paper is organized as follows. In Sec. II we review the Matrix model equations in general supergravity background obtained in \cite{mM0=PRL,mM0=PRD}. There we begin by avoiding the details of the worldline superfield description characteristic for superembedding approach, but rather describing the component worldline equations.
Then, however, we see that some information on the worldline superfield formalism is  necessary to specify the general equations for particular supergravity background, including for purely bosonic ones. Also the supergravity fields necessarily enter the general mM$0$ equations through the pull--back of superfields projected with the use of moving frame and spinor moving frame adapted to the center of energy motion of the mM$0$ system. As a result, an important part of the present paper is devoted to the superfield formalism.

Section III contains some necessary details on supersymmetric pp-wave solutions of 11D supergravity (sec. IIIA) and on its superfield description (sec. IIIB) by a particular 11D superspace $\Sigma^{(11|32)}_{pp-w}$. Sec. IVA contains a brief review of superembedding approach to M0-brane in general 11D supergravity background.  In Sec. IVB we construct a particular worldline superspace ${\cal W}_0^{(1|16)}$ describing the ground state motion of the center of energy of the mM$0$ system, for which all the associated bosonic and fermionic Goldstone fields are equal to zero. We use that in Sec. IVC to obtain the equations of relative motion for the mM$0$ constituents in 11D pp-wave background  for the case when the center of energy motion is described by the above mentioned ground state configuration. These equations coincide with the BMN equations up to the fact that they are obtained for traceless, $su(N)$ valued matrices, rather then for $u(N)$ valued matrix fields like in the case of BMN model. Then we show  that  the complete set of the  BMN equations describes the general motion  of  mM$0$ system in pp-wave background in the leading order on the center of energy Goldstone fields. Particularly, we show in Sec. IVD that the equations for Goldstone fields which follows from the superembedding approach to mM$0$ system coincide with the trace part of the BMN equations. Sec. V contains conclusions and discussion.

\section{Matrix model equations in an arbitrary supergravity background}

\subsection{Equations for matrix fields}

The Matrix model equations in an arbitrary supergravity background were obtained in \cite{mM0=PRL,mM0=PRD} as equations for multiple M$0$-brane (mM$0$) system in an arbitrary 11D supergravity superspace by using superembedding description of mM$0$ proposed in \cite{mM0=PLB}. The set of these equations splits naturally on two subsets:  first describing the center of energy motion and  second describing the relative motion of the constituents of mM$0$ system. The latter subset contains the equations for matrix fields, namely, the {\it traceless} $N\times N$ matrix fields ${\bf  X}^i(\tau)$ and $\Psi_q (\tau)$ depending on one proper time variable $\tau$. 

Here $N$ is the number of M$0$-branes forming the mM$0$ system,
${\bf  X}^i$ is bosonic, takes values in the vector representation of $SO(9)$, so that $i=1,...,9$, and carries  the $SO(1,1)$ wait $2$, ${\bf  X}^i= {\bf  X}_{\#}^i:= {\bf  X}_{++}^i$, while $\Psi_q$
is the fermionic, takes values in the spinor representation of $SO(9)$, so that  $q=1,...,16$, and carries the $SO(1,1)$ weight $3$,  $\Psi_q= \Psi_{\# \,+q}:= \Psi_{++ \,+q}$. 

Already at this stage one can guess that in our discussion we will use the nine-dimensional Dirac matrices $\gamma^i_{qp}$. These are real, symmetric $\gamma^i_{qp}=\gamma^i_{(qp)}$ and obey $\gamma^i\gamma^j + \gamma^j \gamma^i= \delta^{ij} \, I_{16\times 16}$.

The set of the equations of relative motion of mM$0$ constituents contains  the fermionic equation
\begin{widetext}
\begin{eqnarray}\label{M0:DtPsi=}
& D_{\#}{\Psi}_{q}=- {1\over 4} \gamma^i_{qp} \left[ {\bf X}^i\, , \,
\Psi_{p} \right]   + {1\over 24}  \hat{F}_{\# ijk} \gamma^{ijk}_{qr}\Psi_{r}
 - {1\over 4}  {\bf X}^i\hat{T}_{\# i \, +q}\, ,
\end{eqnarray}
the Gauss constraint
\begin{eqnarray}\label{M0:XiDXi=}
& \left[ {\bf X}^i\, , \, D_{\#}{\bf X}^i\,  \right] = 4i \left\{\Psi_{q}\, , \, \Psi_{q} \right\}  \;  \qquad
\end{eqnarray}
and the bosonic equation
\begin{eqnarray}\label{M0:DDXi=}
D_{\#}D_{\#}{\bf X}^i &=
{1\over 16} \left[ {\bf X}^j , \left[ {\bf X}^j ,  {\bf X}^i\,  \right]\right]+ i\gamma^i_{qp} \left\{\Psi_{q}\, , \, \Psi_{p} \right\} +   {1\over 4}  {\bf X}^j \hat{R}_{\# j\; \# i} +
{1\over 8} \hat{F}_{\# ijk} \left[ {\bf X}^j , \, {\bf X}^k\,  \right] -2i \Psi_{q}\hat{T}_{\# i \, +q}  \, . \;
\end{eqnarray}
\end{widetext}
The left hand sides ({\it l.h.s.}'s) of these equations involve   the covariant time derivative $D_{\#}$  which we describe below (see sec. IIC); it contains
derivative with respect to the proper time variable $\tau$
($D_{\#}={\cal E}_{\#}^\tau(\tau) \partial_\tau +...$), but in general case other contributions are also present.

The {\it r.h.s.}'s  of Eqs. (\ref{M0:DtPsi=}) and (\ref{M0:DDXi=}) involve the following projections of the pull--backs of the superfield supergravity 'fluxes' ({\it i.e.} of the field strengths superfields)
\begin{eqnarray} \label{M0:Fluxes}
  \hat{F}_{\# ijk}&:=& F^{{a}{b}{c}{d}} (\hat{Z}) u_{{a}}
{}^{=}u_{{b}}{}^{i}u_{{c}}{}^{j}u_{{d}}{}^{k}\; , \qquad    \\  \label{M0:RFlux} \hat{R}_{\# ij \#} &:= & R_{dc\;
ba}(\hat{Z})u^{d=}u^{ci} u^{bj}u^{a=} \; ,  \qquad
\\ \label{M0:fFluxe}  \hat{T}_{\#\, i\, +q}
&:= & T_{{a}{b}}{}^{\alpha}(\hat{Z})\,  v_{{\alpha}q}^{\; -}\,
u_{{a}}^{=}u_{{b}}^i\; .    \qquad
\end{eqnarray}
Here $F^{{a}{b}{c}{d}} =F^{[{a}{b}{c}{d}]} ({Z})$, $T_{ab}{}^\alpha= T_{[ab]}{}^\alpha (Z)$ and $R_{dc\; ba}=R_{[dc]\;[ba]}(Z)$ are superfield generalizations of the  field strength  of the 3-rd rank antisymmetric tensor gauge field, of the gravitino field strength and of the Riemann tensor of the 11D supergravity. We discuss their properties in the next subsection \ref{secFlux}. In Eqs. (\ref{M0:Fluxes}), (\ref{M0:RFlux}) and (\ref{M0:fFluxe}) these flux superfields depend on bosonic and fermionic coordinate functions $\hat{Z}^{{ {M}}}(\tau)=
(\hat{x}{}^{ {m}}(\tau)\, , \hat{\theta}^{\check{ {\alpha}}}(\tau))$ which define the embedding of the worldline $W^1$, on which the matrix fields ${\bf X}^i(\tau)$ and $\Psi_q(\tau)$ are defined, in the target 11D superspace,
\begin{eqnarray}
\label{WinSt} & {W }^{1}\in \Sigma^{(11|32)} : \quad
Z^{ {M}}=  \hat{Z}^{ {M}}(\tau ) =
(\hat{x}{}^{ {m}}(\tau), \hat{\theta}^{\check{ {\alpha}}}(\tau))  . \qquad
\end{eqnarray}
This worldline $W^1$ is naturally associated with the movement of the center of energy of the mM$0$ system so that its local coordinate $\tau$,  can be called {\it center of energy proper time}. The center of energy motion  resembles  a motion of a particle or a single $0$-brane (we will discuss this below). Then, using the experience of studying single (super)particle or (super-)p-brane we can state that the (not pure gauge part of the) {\it coordinate functions $\hat{Z}^{{ {M}}}(\tau)=
(\hat{x}{}^{ {m}}(\tau)\, , \hat{\theta}^{\check{ {\alpha}}}(\tau))$ are bosonic and fermionic Goldstone fields corresponding to the translation symmetry and supersymmetry which are broken spontaneously} due to the presence of the the effective worldline $W^1$ describing the {\it center of energy} motion of the mM0 system.

Furthermore, as  ${\bf D}_\#$ in Eqs.  (\ref{M0:DtPsi=}), (\ref{M0:XiDXi=}) and (\ref{M0:DDXi=}) is a covariant time derivative on  $W^1$ defined in (\ref{WinSt}), in general case it contains the contributions from the coordinate functions $\hat{x}{}^{ {m}}(\tau)$ and $\hat{\theta}^{\check{ {\alpha}}}(\tau)$ (see subsection \ref{secD++}).

Finally, the worldline fields $u_{{a}}{}^{=}=u_{{a}}{}^{=}(\tau)$, $u_{{b}}{}^{i}=u_{{b}}{}^{i}(\tau)$ and $v_{{\alpha}q}^{\; -}=v_{{\alpha}q}^{\; -}(\tau)$, which are used in Eqs. (\ref{M0:Fluxes}), (\ref{M0:RFlux}) and (\ref{M0:fFluxe}), define the moving frame adapted to the center of energy motion of the mM$0$ system and its spinorial representation (spinor moving frame). There indices
$a=0,1,...,9, 10$ and $\alpha = 1,...,32$ correspond to the vector and spinor representations of the eleven-dimensional Lorentz group $SO(1,10)$, while $-$ denotes the scaling dimension ($-1$ for $v_{{\alpha}q}^{\; -}(\tau)$ and $-2$ for
$u_{{a}}{}^{=}:=u_{{a}}{}^{--}$)  with respect to $SO(1,1)$ subgroup of $SO(1,10)$, and $i=1,...,9$ and $q=1,...,16$ are vector and spinor indices of $SO(9)\subset SO(1,10)$. 

As we discuss below, these moving frame variables can be used (instead of the coordinate functions (\ref{WinS}) or together with these) to describe the center of energy motion of the mM$0$ system. It is useful to keep in mind the simplest case, when the frame is not actually moving, which is described by constant  $u_{{a}}{}^{= }=\delta_a^0-\delta_a^{10}$,  $u_{{a}}{}^{i}=\delta_a^i$ and $v_{{\alpha}q}^{\; -}= \delta_{\alpha}^{-q}$. Actually this simplest frame is sufficient for the major part of our study here; only in the second part of sec. IVB we use a more complicated moving frame.

\subsection{Flux superfields (superfield generalizations of the field strength) of D=11 supergravity}
\label{secFlux}

The flux superfields $F^{{a}{b}{c}{d}} =F^{[{a}{b}{c}{d}]} ({Z})$, $T_{ab}{}^\alpha= T_{[ab]}{}^\alpha (Z)$ and $R_{dc\;
ba}=R_{[dc]\;[ba]}(Z)$, entering Eqs. (\ref{M0:Fluxes}), (\ref{M0:RFlux}) and (\ref{M0:fFluxe}), satisfy the superfield generalization of the 11D supergravity equations of motion, the set of which includes  Einstein equations
\begin{eqnarray} \label{Eq=Einstein}
R_{ab}= - {1\over 3}F_{ac_1c_2c_3}F_{b}{}^{c_1c_2c_3}  +  {1\over 36}\eta_{ab} F_{c_1c_2c_3c_4}F^{c_1c_2c_3c_4}\; , \nonumber \\ \qquad R_{ab}:= R_{ac\, b}{}^c\; , \qquad \eta_{ab}=diag(+,-,\ldots , -)\;  \qquad
\end{eqnarray}
and the Rarita-Schwinger equations $T_{bc}{}^\beta \Gamma^{abc}_{\beta\alpha}=0$. It is convenient to write this latter in the equivalent form of 
\begin{eqnarray} \label{Eq=RS}
T_{ab}{}^\beta \Gamma^{b}_{\beta\alpha}=0\; . \qquad
\end{eqnarray}
Here and below $\Gamma^{b}_{\alpha\beta}= \Gamma^{b}_{\beta\alpha}= \Gamma^{b}{}_{\alpha}{}^\gamma C_{\gamma\beta}$,  where  $\Gamma^{b}{}_{\alpha}{}^\beta$   are eleven dimensional Dirac matrices obeying $\Gamma^{(a} \Gamma^{b)}:=  {1\over 2}(\Gamma^{a}\Gamma^{b}+\Gamma^{b}\Gamma^{a})= \eta^{ab}\, I_{32\times 32}$, and 
$C_{\beta\gamma}=- C_{\gamma\beta}$ is the 11D charge conjugation matrix, $\Gamma^{ab}=\Gamma^{[a} \Gamma^{b]}:=  {1\over 2}(\Gamma^{a}\Gamma^{b}-\Gamma^{b}\Gamma^{a})$, $\Gamma^{abc}=\Gamma^{[a} \Gamma^{b}\Gamma^{c]}$, etc. 
See {\it e.g.}  \cite{IB09:M-D,mM0=PRD} for  more details on the properties and the explicit representations of the 11D $\Gamma$ matrices.

As it was shown in \cite{CremmerFerrara80,BrinkHowe80}, the terms of higher order in the decomposition of these superfields on Grassmann coordinates are expressed through their leading components and supergravity potentials so that no new degrees of freedom appear \cite{CremmerFerrara80,BrinkHowe80} (see also \cite{BdAPV05} reviewing this in the present notation).
In particular,
\begin{eqnarray}
\label{DF4=TG} D_{\alpha}F_{abcd}=- 6\, T_{[ab}{}^\beta \Gamma_{cd]}{}_{\beta\alpha}\; , \qquad
\\
\label{DTabf=RG+Dt+tt} D_{\alpha}T_{ab}{}^{\beta}=-{1\over 4} R_{ab}{}^{cd} \Gamma_{cd}{}_{\alpha}{}^{\beta} - 2(D_{[a}t_{b]} +t_{[a}t_{b]})_{\alpha}{}^{\beta}\; , \qquad \end{eqnarray}
where $t_{a\alpha}{}^\beta$ is expressed through $F_{abcd}(Z)$ by
\begin{eqnarray}\label{ta:=}
t_{a\beta}{}^{\!\alpha}&:= {i \over 18}
\left(F_{abcd}\Gamma^{bcd}{}_\beta^{\;\alpha} + {1\over 8}
F^{bcde} \Gamma_{abcde}{}_\beta^{\;\alpha}\right) \; . \qquad
\end{eqnarray}

The pull--backs of the superfields in Eqs. (\ref{M0:Fluxes}), (\ref{M0:RFlux}) and (\ref{M0:fFluxe}) are obtained by substituting the center of energy coordinate functions $\hat{Z}^{{ {M}}}(\tau)=
(\hat{x}{}^{ {m}}(\tau)\, , \hat{\theta}^{\check{ {\alpha}}}(\tau))$ for superspace coordinates ${Z}^{{ {M}}}=
({x}{}^{ {m}}\, , {\theta}^{\check{ {\alpha}}})$. As a result, in general case, the {\it r.h.s.}'s of Eqs. (\ref{M0:DtPsi=}) and (\ref{M0:DDXi=}) contain, besides $\hat{x}{}^{ {m}}(\tau)$, also the contributions from the fermionic coordinate functions  $\hat{\theta}^{\check{ {\alpha}}}(\tau)$. In particular, taking into account (\ref{DF4=TG}) and using the Wess--Zumino gauge for superfield supergravity one finds that
\begin{eqnarray}
\label{hF4=hF4x+}
\hat{F}_{abcd}= F_{abcd}(\hat{x}(\tau))- 6\hat{\theta}{}^\alpha (\tau)  \Gamma_{[ab|}{}_{\alpha\beta}T_{|cd]}{}^\beta (\hat{x}(\tau)) + \nonumber \\ + \propto \hat{\theta}{}^\alpha (\tau)  \hat{\theta}{}^\beta (\tau)
\;  . \qquad
\end{eqnarray}
Thus, generically the structure of Eqs. (\ref{M0:DtPsi=}), (\ref{M0:DDXi=}) and (\ref{M0:XiDXi=}) seems to be quite complicated. It simplifies for the target supergravity superspace describing the purely bosonic 
supergravity solutions described by superspaces with vanishing gravitino field strength superfield, $T_{ab}{}^\alpha(Z)=0$.
The expression for pull--back of bosonic fluxes entering the {\it r.h.s.}'s of Eqs. (\ref{M0:DtPsi=}) and (\ref{M0:DDXi=}) in these cases do not contain the center of energy Goldstone fermion contributions (for instance (\ref{hF4=hF4x+}) reduces to   $\hat{F}_{abcd}= F_{abcd}(\hat{x}(\tau))$). 
The completely supersymmetric $AdS_4\otimes S^7$, $AdS_4\otimes S^7$ and pp-wave superspaces are examples of such backgrounds characterized by constant and covariantly constant 4-form field strength $F_{abcd}$ and Riemann tensor $R_{cd}{}^{ab}$. 

However, even for the  cases of purely bosonic supergravity solutions the structure of $D_\#$ is not so simple.

\subsection{Covariant derivative $D_\#$}
\label{secD++}

The equations of the relative motion of mM$0$ constituents, Eqs.  (\ref{M0:DtPsi=}), (\ref{M0:XiDXi=}), (\ref{M0:DDXi=}), are obtained in the frame of superembedding approach, which provides a superfield description of the dynamics of mM$0$ system. This implies that the  matrix fields are leading components of some matrix superfields  \begin{eqnarray}\label{M0:X=X+eta}
{\bf X}^i(\tau , \eta )= {\bf X}^i(\tau) + \propto \eta^{\check{p}}\; , \qquad \nonumber \\  {\Psi}_{q}(\tau , \eta )= {\Psi}_{q}(\tau) + \propto \eta^{\check{p}}\; , \qquad
\end{eqnarray}
which we denote by the same symbol.  These depend not only on the bosonic variable $\tau$, but also on $16$ fermionic variables   $\eta^{\check{p}}$ (obeying $\eta^{\check{q}}\eta^{\check{p}}=-\eta^{\check{p}}\eta^{\check{q}}$)
), this is to say, they are functions on a (generically curved) superspace ${\cal W}^{(1|16)}$ with one bosonic and 16 fermionic directions,
\begin{eqnarray}\label{M0:W1-16=teta}
{\cal W}^{(1|16)} \; = \; \{\zeta^{{\cal M}} \}= \{(\tau , \eta^{\check{q}}) \} \; , \qquad \\ \nonumber  \eta^{\check{q}}\eta^{\check{p}}=-\eta^{\check{p}}\eta^{\check{q}}\, ,  \quad {\check{q}}, {\check{p}}=1,...,16\; . 
\end{eqnarray}
 This superspace can be associated with the center of energy motion of the mM$0$ system and can be called {\it center of energy worldline superspace}.
The  covariant derivative $D_{\#}$, which enters Eqs.  (\ref{M0:DtPsi=}), (\ref{M0:XiDXi=}), (\ref{M0:DDXi=}) is defined as a leading component of the bosonic covariant derivative of ${\cal W}^{(1|16)}$,
\begin{eqnarray}\label{M0:DXi:=}
D_{\#}{\bf X}^i(\tau) = \left({\cal D}_{\#}{\bf X}^i(\tau , \eta )\right)\vert_{\eta^{\check{q}}=0} \, , \qquad \nonumber
\\ D_{\#}{\Psi}_{q}(\tau) = \left({\cal D}_{\#}{\Psi}_{q}(\tau , \eta )\right)\vert_{\eta^{\check{q}}=0} \, . \qquad
\end{eqnarray}
As far as this latter appears in the decomposition of the covariant differential ${\cal D}$ on the supervielbein of ${\cal W}^{(1|16)}$,
\begin{eqnarray}\label{cE=cEcE}
{\cal E}^{{\cal A}}= ({\cal E}^{\#},  {\cal E}^{+q})= d\tau  {\cal E}_\tau^{{\cal A}}(\tau, \eta)+d\eta^{\check{p}}  {\cal E}_{\check{p}}^{{\cal A}}(\tau, \eta)\; , \qquad
\\
\label{cD=W1-16}
{\cal D}:= {\cal E}^{\#} {\cal D}_{\#}+ {\cal E}^{+q} {\cal D}_{+q}:=
d{\tau} {\cal D}_{\tau}+ d{\eta}^{\check{q}} {\cal D}_{\check{q}}\, , \qquad \\
\label{d=W1-16}
d =
d{\tau} \partial_{\tau}+ d{\eta}^{\check{q}} \partial_{\check{q}}=
d{\tau} {\partial\over \partial \tau}+ d{\eta}^{\check{q}}  {\partial\over \partial {\eta}^{\check{q}} }
, \qquad
\end{eqnarray}
the $D_\#$ in Eqs.  (\ref{M0:DtPsi=}), (\ref{M0:XiDXi=}), (\ref{M0:DDXi=}) contains, besides the term with  $\partial_\tau $ and $SO(1,1)\times SO(9)\times SU(N)$ connection, also the contribution from the 'worldline gravitino' ${\cal E}_\tau^{+q}$,
\begin{eqnarray}\label{D++=b+f}
D_{\#}  =  {\cal E}_\#^\tau (\tau) ({\cal D}_{\tau}...)\vert_{\eta=0} + {\cal E}_{\#}^{\check{q}}(\tau) ({\cal D}_{\check{q}}...)\vert_{\eta=0} = \nonumber \qquad \\
= ({\cal E}_\#^\tau (\tau) + {\cal E}_{\#}^{\check{q}}(\tau){\cal E}_{\check{q}}^\tau (\tau)) ({\cal D}_{\tau}...)\vert_{\eta=0} +\nonumber \qquad \\
+ {\cal E}_{\#}^{\check{q}}(\tau){\cal E}_{\check{q}}^{+p}(\tau)) ({\cal D}_{+p}...)\vert_{\eta=0} \; .
 \qquad
\end{eqnarray}
Here
\begin{eqnarray}\label{cE-1}
{\cal E}_{\#}^{\check{q}}(\tau)= - {\cal E}_{\tau}^{+{p}}{\cal E}_{+p}{}^{\check{q}}/{\cal E}^{\#}_\tau\;  \qquad
\end{eqnarray}
and ${\cal E}_{\#}^{\tau}=(1 - {\cal E}_{\tau}^{+{p}}{\cal E}_{+p}{}^{\tau})/{\cal E}^{\#}_\tau$ as well as ${\cal E}_{+p}{}^{\tau}$ and ${\cal E}_{+p}{}^{\check{q}}$ are elements of the inverse supervielbein matrix.

As it was shown in \cite{mM0=PLB}, the fermionic covariant derivative of the bosonic matrix superfield ${\bf X}^i$ is expressed through the fermionic superfield $\Psi_q$ by
\begin{eqnarray}\label{M0:DX=gP}
D_{+q}{\bf X}{}^i= 4i\gamma^i_{qp}\Psi_{q}\;  \qquad
\end{eqnarray}
so that $D_{\#}{\bf X}{}^i $ entering  Eqs.   (\ref{M0:XiDXi=}) and (\ref{M0:DDXi=}) reads 
\begin{eqnarray}\label{D++Xi=+EgP}
D_{\#}{\bf X}{}^i = ({\cal E}_\#^\tau (\tau) + {\cal E}_{\#}^{\check{q}}(\tau){\cal E}_{\check{q}}^\tau (\tau)) {\cal D}_{\tau} {\bf X}{}^i(\tau)  - \nonumber \qquad \\ -4i
 {\cal E}_{\tau}^{+{p}}{\cal E}_{+p}{}^{\check{q}^\prime }{\cal E}_{\check{q}^\prime}^{+{q}}(\tau)
\, \gamma^i_{qp}\Psi_{p}(\tau)/{\cal E}^{\#}_\tau \;  \qquad
\;   \qquad
\end{eqnarray}
and contains, besides  $\propto {\cal D}_{\tau} {\bf X}{}^i(\tau) $ term, the contribution $\propto \Psi_{p}(\tau)$.   
Similarly, the expression for $D_\# \Psi_q$ in Eq.  (\ref{M0:DtPsi=}) contains, besides $D_\tau \Psi_p$, also 
 $\propto  {\bf X}{}^i(\tau)$ contributions\footnote{As shown in \cite{mM0=PRL, mM0=PRD},   
$D_{+p}\Psi_{q} = {1\over 2}\gamma^i_{pq}D_{\#}{\bf X}^i +
{1\over 16} \gamma^{ij}_{pq} [{\bf X}^i, {\bf X}^j]- {1\over 12} {\bf
X}^i\hat{F}_{\# jkl}\left(\delta^{i[j}\gamma^{kl]}+ {1\over 6}
\gamma^{ijkl}\right)_{pq}
 \; . $
}

The expression for ${\cal D}_{+p}\Psi_q$, which can be found in \cite{mM0=PRL, mM0=PRD}, then so is the expression for $D_\# \Psi_q$ in Eq.  (\ref{M0:DtPsi=}).

Notice that the expressions for $D_{\#}{\bf X}{}^i$ and  $D_\# \Psi_q$ simplify essentially 
when the worldline gravitino vanishes,
\begin{eqnarray}\label{D++Xi=simple}
 {\cal E}_{\tau}^{+{p}}=0 \quad \Rightarrow \quad \begin{cases}
 D_{\#}{\bf X}{}^i = {\cal E}_\#^\tau (\tau)  {\cal D}_{\tau} {\bf X}{}^i(\tau)  \cr   D_{\#}\Psi_{q} = {\cal E}_\#^\tau (\tau)  {\cal D}_{\tau} \Psi_{q}(\tau)\; . \end{cases}
\end{eqnarray}
As we show below, this happens for the mM$0$ system in pp-wave superspace when the center of energy Goldstone fermion is set to zero.

The general conclusion which we have caught  in this and previous  subsections is that some knowledge on the worldline superspace formalism is still necessary to extract the Matrix model equation in a particular 11D supergravity background from the equations of motion of mM$0$ system in an arbitrary supergravity superspace obtained in \cite{mM0=PRL,mM0=PRD}.

We will also need to describe the center of energy dynamics of mM$0$ system in pp-wave background. In the next section we review the equations of the  mM$0$ center of energy motion in general supergravity background proposed in \cite{mM0=PRL,mM0=PRD}.

\subsection{Equations of the center of energy motion, moving frame and spinor moving frame}

It is natural to formulate the equations of center of energy motion of the mM$0$ system in terms of the coordinate functions $\hat{Z}^{ {M}}(\tau ) =
(\hat{x}{}^{ {m}}(\tau)\, , \hat{\theta}^{\check{ {\alpha}}}(\tau))$ of Eq. (\ref{WinSt}).
However, it can be also described with the use of {\it moving frame variables} which can be considered as elements of the $SO(1,10)$ valued moving frame matrix
 \begin{eqnarray}\label{harmUin}
U_b^{(a)}= \left({u_b^{=}+ u_b^{\#}\over 2}, u_b^{i}, {
u_b^{\#}-u_b^{=}\over 2} \right)\; \in \; SO(1,10)\; , \quad  \\ a,b=0,1,...,9, 10\; , \quad i=1,...,9\, .  \nonumber
\end{eqnarray}
The above statement of Lorentz group valuedness of the moving frame matrix is tantamount to saying that the moving frame vectors obey the constraints \cite{Sok}
\begin{eqnarray}\label{u--u--=0}
u_{ {a}}^{=} u^{ {a}\; =}=0\; , \quad    u_{ {a}}^{=} u^{ {a}\,i}=0\; , \qquad u_{ {a}}^{\; = } u^{ {a} \#}= 2\; , \qquad
 \\  \label{u++u++=0} u_{ {a}}^{\# } u^{ {a} \#
}=0 \; , \qquad
 u_{{a}}^{\;\#} u^{ {a} i}=0\; , \qquad  \\  \label{uiuj=-} u_{ {a}}^{ i} u^{ {a} j}=-\delta^{ij}\; .  \qquad
\end{eqnarray}
Notice that the light-like moving frame vector $u_b^{=}$  has already appeared in Eqs. (\ref{M0:fFluxe}), (\ref{M0:DfFluxe}), (\ref{M0:RFlux}).

In massless superparticle model the moving frame variables appear as a counterpart of the momentum (see \cite{IB07:M0} and refs therein) in the sense that the pull--back of the bosonic supervielbein form to the worldline is written as
\begin{eqnarray}\label{hEa=}
 \hat{E}{}^a:= d\tau \hat{E}{}_\tau^a:= d\hat{Z}^{\underline{M}}(\tau) E_{\underline{M}}{}^a(\hat{Z}) = {1\over 2}{\cal E}^\# u^{= a}   \; ,  \quad \\ \nonumber \Leftrightarrow \quad \hat{E}{}_\tau^a:= \partial_\tau \hat{Z}^{\underline{M}}(\tau) E_{\underline{M}}{}^a(\hat{Z}) = {1\over 2}{\cal E}_\tau^\# (\tau) u^{= a}(\tau)\; ,
\end{eqnarray}
were ${\cal E}^\#(\tau)=d\tau\, {\cal E}_\tau^\# (\tau)$ is the einbein form on $W^1$.

Indeed, Eq. (\ref{hEa=}) implies that
\begin{eqnarray}\label{M0:u--=} & u^{= a} = 2\hat{E}{}^a_\# =2\partial_\tau{\hat{X}}/{\cal E}_\tau^\# + \ldots \; , \qquad
\end{eqnarray}  so that it is not surprising that the
bosonic equations of the center of energy motion of the mM$0$ system are formulated in terms of $1/2 D_\# u^{= a}= D_\#\hat{E}{}^a_\#= \partial_\tau \partial_\tau {\hat{X}}/
({\cal E}_\tau^\#)^2 + \ldots $.

In particular, the bosonic equations of the mM$0$ center of energy motion obtained in \cite{mM0=PLB} coincide with the equation of motion of a single M$0$-brane and imply
\begin{eqnarray}\label{M0:bEq}
D_{\# } u^{=  {a}}\, =
2D_{\# } \hat{E}_{\# }{}^{  {a}} =0 \; . \qquad
\end{eqnarray}

To write the fermionic equations of the center of energy motion, we need to use
the spinor moving frame variables (or spinorial Lorentz harmonics
\cite{B90,Ghsds,BZ-str,GHT93}, see  \cite{IB07:M0} for more references). These can be defined as elements of the $Spin(1,10)$ valued matrix \begin{eqnarray}\label{harmVin}
V_{(\beta)}^{\;\;\; \alpha}= \left(\begin{matrix}  v^{+\alpha}_q
 \cr  v^{-\alpha}_q \end{matrix} \right) \in Spin(1,10)\; 
 \;  \qquad
\end{eqnarray}
which is a double covering of the moving frame matrix. This latter statement implies, in particular, that $v_q^{-\alpha}=v_q^{-\alpha}(\tau)$ can be considered  as square root of the light--like vector $u^{=}_{ a}$ in the sense of that it obeys the following constraint
\begin{eqnarray} \label{Iu--=vGv}  \delta_{qp}
u^{=}_{ a}= v_q^{-\alpha} \Gamma^a_{\alpha\beta} v_p^{-\beta}\; .
\qquad
\end{eqnarray}
The fermionic equations of the center of energy motion of the mM$0$ system, which in the superembedding approach of \cite{mM0=PLB,mM0=PRL,mM0=PRD} coincide with the fermionic equation of motion of a single M$0$-brane, can be written as
\begin{eqnarray}\label{M0:DiracEq}
 \hat{{E}}_{\# }{}^{ {\alpha}}v_{ {\alpha}p}{}^-
=0 \;   \qquad  \end{eqnarray}
($\hat{{E}}_{\# }{}^{ {\alpha}}= {\cal E}_{\# }{}^\tau \partial_\tau  \hat{{\theta}}{}^{ {\alpha}}+\ldots$). To make their form more standard, one can use Eqs. (\ref{Iu--=vGv}) and (\ref{M0:u--=}) to show that (\ref{M0:DiracEq}) implies
\begin{eqnarray}\label{M0:DiracEq'}
\hat{{E}}_{\# }{}^{{a}} \, \Gamma_{a {\alpha}{\beta}}\hat{{E}}_{\# }{}^{ {\beta}}
=0 \;  . \qquad  \end{eqnarray}
When the worldline gravitino vanishes,  this is equivalent to
\begin{eqnarray}\label{M0:DiracEq''}
\hat{{E}}_{\tau }{}^{{a}} \, \Gamma_{a {\alpha}{\beta}}\hat{{E}}_{\tau  }{}^{ {\beta}}
=0 \;  . \qquad  \end{eqnarray}

More details on the  moving frame and spinor moving frame variables can be found in \cite{mM0=PRD} and in refs. therein. Here we will need to know only a few of their properties, in particular that, on the  shell of the equations of the mM$0$ center of energy motion the $SO(1,10)\times SO(1,1)\times SO(9)$ covariant derivatives of $u^{i}_a$ is expressed in terms of $u^{=}_{ a}$ and the covariant derivative of $u^{=}_{a}$ vanishes. Actually
\begin{eqnarray}
 \label{Dui}  &  Du^{i}_a
 =
 {1\over 2} u^{=}_a\Omega^{\# i}  , \quad Du^{=}_a= 0 \, , \quad Du^{\# }_a= u^{i}_a \Omega^{\# i}   , \quad
\end{eqnarray}
with the same one form coefficient $\Omega^{\# i}$ from $ Du^{i}_a$ enter the expression for $ Du^{\#}_a$ and
$Dv_q^{+\alpha}$, while $v_q^{-\alpha}$ is covariantly constant,
\begin{eqnarray}
\label{Dv-q=0} &  Dv_q^{-\alpha} = 0\; , \qquad  Dv_q^{+\alpha}  = - {1\over 2} \Omega^{\# i}
v_p^{-\alpha} \gamma_{pq}^{i}\; . \qquad
\end{eqnarray}
Notice that the above relation also define the $SO(1,1)\times SO(9)$ connection induced by (super)embedding and that their worldline superfield generalizations are also valid, see \cite{mM0=PLB,mM0=PRD}.

\subsection{Fluxes allowing  center of energy motion preserving 1/2 of the target space supersymmetry}

An important observation is that with the above consequences of the center of energy equations of motion, the derivative acting on the projections of the pull--backs of fluxes of Eqs.  (\ref{M0:Fluxes}), (\ref{M0:RFlux}) and (\ref{M0:fFluxe}) reduce to the projections of the pull--backs of the targets superspace derivatives of the fluxes,
\begin{eqnarray} \label{M0:DFluxes}
  D\hat{F}_{\# ijk}&:=&  u_{{a}}
{}^{=}u_{{b}}{}^{i}u_{{c}}{}^{j}u_{{d}}{}^{k}\, DF^{{a}{b}{c}{d}}(Z)\vert_{Z=\hat{Z}}\; , \quad   \nonumber \\  D\hat{R}_{\# ij \#} &:= & u^{d=}u^{ci} u^{bj}u^{a=}\, DR_{dc\;
ba}(Z)\vert_{Z=\hat{Z}} \; ,  \quad \nonumber
\\ \label{M0:DfFluxe}  D\hat{T}_{\#\, i\, +q}
&:= &  v_{{\alpha}q}^{\; -}\,
u_{{a}}^{=}u_{{b}}^i \, DT_{{a}{b}}{}^{\alpha}({Z})\vert_{Z=\hat{Z}}\; .    \quad
\end{eqnarray}
Then, the above mentioned fact that the superfield generalizations of Eqs. (\ref{Dui}), (\ref{Dv-q=0}) are  also valid allows \cite{mM0=PRD} to deduce the supersymmetry transformations of the projections of the pull--backs of the fluxes from the superfield supersymmetry of target superspace.
In particular, as it was found in \cite{mM0=PRD}, the center of energy motion of mM$0$ system in  superspaces describing purely bosonic solutions of supergravity ({\it i.e.} superspaces with $T_{{a}{b}}{}^{\alpha}=0$) can preserve 1/2 of the target (super)space supersymmetry if the projections of the pull--backs of the bosonic fluxes to the center of energy worldline (superspace) obey  (a worldline superfield generalization of) the following relations 
\footnote{In \cite{mM0=PRD} there are misprints in the coefficients for the second and third term in the counterpart of Eq. (\ref{DpTq=0->R}) (Eq. (5.10) of \cite{mM0=PRD}); the correct values, which coincide with the ones in Eq. (\ref{DpTq=0->R}) above, can be seen from Eq. (5.7) of \cite{mM0=PRD}.}
 \begin{eqnarray} \label{DpTq=0->R}
 \hat{R}_{\# ij\#} - {1\over 6}\hat{F}_{\# ikl}\hat{F}_{\# klj} - {1\over 54}\delta^{ij} (\hat{F}_{\# k_1k_2k_3})^2=0 \; , \\ \label{DpTq=0->DF}
 D_{\#} \hat{F}_{\# ijk} =0 \; , \qquad \\ \label{DpTq=0->FF3}
 \hat{F}_{\# ij[k_1}\hat{F}_{\# k_2k_3]j}=0\; .\qquad  \end{eqnarray}

Notice also that the projection (\ref{M0:RFlux}) of the Riemann tensor is symmetric (due to the Bianchi identities  $R_{[abc]d}=0$). Furthermore, its trace (on SO(9) vector indices)  is expressed through the product of the projections (\ref{M0:Fluxes}) of the 4-form fluxes by
\begin{eqnarray}\label{hatEiEq=singlet}
\hat{R}_{\# j\# j} + {1\over 3} (\hat{F}_{\# ijk})^2= 0 \; , \qquad
\end{eqnarray}
which is the $u^{=}_au^{= b}$ projection of the pull--back of
the supergravity Einstein equation (\ref{Eq=Einstein}) to ${\cal W}^{1}$.

In the remaining part of this paper we will consider mM$0$ system in the maximally supersymmetric 11D pp-wave background.

\section{supersymmetric 11D pp-wave solution and its superfield description}
\setcounter{equation}0
\subsection{Supersymmetric pp-wave solution of the 11D supergravity equations}

The completely supersymmetric pp-wave solution of $D=11$ supergravity equations is characterized by
the 11D spacetime interval
\begin{widetext}
\begin{eqnarray}\label{ppw=ds2}
ds^2= dx^{--}dx^{++}+ \left[ \left({\mu\over 3}\right)^2 x^Ix^I + \left({\mu\over 6}\right)^2 x^{\tilde{J}} x^{\tilde{J}} \right]dx^{++}dx^{++}- dx^Idx^I - dx^{\tilde{J}} dx^{\tilde{J}}\;  , \qquad \begin{cases} I=1,2,3, \cr {\tilde{J}}= 4,5,6,7,8,9, \end{cases}
\end{eqnarray}
\end{widetext}
and by the constant 4 form flux \footnote{Our notation here are close to \cite{BMN} where one might also find a number of  references on pp-wave solutions of supergravity equations.}
\begin{eqnarray}\label{pp-w=F}
F_{abcd} = 2\mu \delta_{[a}{}^{++}\delta_{b}^{I}\delta_{c}^{J}\delta_{d]}^{K}\epsilon_{IJK}\; .
 \qquad
 \end{eqnarray}
 The dual seven form flux is then given by
\begin{eqnarray}\label{pp-w=F7}
F^{c_1\ldots c_7}={1\over 4!}\epsilon^{c_1\ldots c_7b_1\ldots b_4}F_{b_1\ldots b_4}= \qquad \nonumber \\ -{7\mu\over 2}\eta^{++[c_1}\delta_{\tilde{J}_1}{}^{c_2}\ldots \delta_{\tilde{J}_6}{}^{c_7]}\epsilon^{\tilde{J}_1\ldots \tilde{J}_6}\; . 
\end{eqnarray}
The corresponding vielbein one-forms and nonvanishing components of the spin connection read
\begin{eqnarray}\label{ppw=e++}
\nonumber  
e^{--}= dx^{--}+ \left[ \left({\mu\over 3}\right)^2 x^Ix^I + \left({\mu\over 6}\right)^2 x^{\tilde{J}} x^{\tilde{J}} \right]dx^{++}\; , \qquad \\  e^{++}=dx^{++}\; , \qquad  e^I=dx^I\; , \qquad e^{\tilde{J}}=dx^{\tilde{J}}\; , \qquad 
\\ \label{ppw=om--i}
\omega^{I--}=
2\left({\mu\over 3}\right)^2 e^{++} x^I = 2\left({\mu\over 3}\right)^2 dx^{++}\, x^I\; , \nonumber  \qquad \\ \omega^{\tilde{J}--}= 2\left({\mu\over 6}\right)^2 e^{++} x^{\tilde{J}}= 2\left({\mu\over 6}\right)^2 dx^{++} \, x^{\tilde{J}} \; . \qquad
\end{eqnarray}
As a result the only nonvanishing components of the $SO(1,10)$ curvature 2--form are
\begin{eqnarray}\label{ppw=R--I}
& R^{--I}= - 2 \left({\mu\over 3}\right)^2\; e^{++}\wedge e^I \; , \qquad  \nonumber \\ & R^{--\tilde{J}} = -2  \left({\mu\over 6}\right)^2 \; e^{++}\wedge e^{\tilde{J}}\; ,  \qquad
\end{eqnarray}
so that the only nonvanishing components of the Riemann curvature tensor 
 are
\begin{eqnarray}\label{ppw=R++++}
& R_{++ IJ ++}\equiv {1\over 4} R^{-- IJ --}= -  \left({\mu\over 3}\right)^2\delta^{IJ}\; , \qquad \nonumber \\ & R_{++ \tilde{I}\tilde{J} ++}\equiv {1\over 4} R^{-- \tilde{I}\tilde{J} --}= - \left({\mu\over 6}\right)^2\delta^{\tilde{I}\tilde{J}}\; , \qquad
\end{eqnarray}
Using (\ref{ppw=R++++}) and (\ref{pp-w=F}) one can easily check that the Einstein equation (\ref{Eq=Einstein}) is satisfied.

\subsection{Supersymmetric pp-wave solution of the 11D superspace supergravity constraints}

The 11D superspace  representing the completely supersymmetric pp-wave solution of the 11D supergravity, which we denote by $\Sigma^{(11|32)}_{pp-w}$, was described {\it e.g.} in \cite{pp-w=DSR}. It can be defined through the following Maurer-Cartan equation (reduction of the 11D supergravity constraints from \cite{CremmerFerrara80,BrinkHowe80})
\begin{eqnarray}\label{Tua=ppw}
T^{\underline{a}}&=&-  i E^\alpha \wedge E^\beta \Gamma^{\underline{a}}_{\alpha\beta} \, ,
\quad
T^\alpha=  - E^{\underline{a}} \wedge E^\beta T_{\underline{a}\, \beta }{}^\alpha \, ,\quad
 \; \\
\label{RL=AdS4} R^{\underline{a}\underline{b}}&=& {1\over 2} E^\alpha \wedge E^\beta
R_{\alpha\beta}{}^{\underline{a}\underline{b}} + {1 \over 2} E^{\underline{d}} \wedge E^{\underline{c}}
R_{\underline{c}\underline{d}}{}^{\underline{a}\underline{b}}\; , \qquad
\end{eqnarray}
with constant $T_{\beta \underline{a}}{}^\alpha$ and $R_{\alpha\beta}{}^{\underline{a}\underline{b}}$ which are constructed from the constant flux (\ref{pp-w=F}) as
\begin{eqnarray}\label{Tbff=}
 T_{{a} \, \beta }{}^\alpha =
 {i \over 18}
\left(F_{{a}[3]}
\Gamma^{[3]} + {1\over 8}
F^{[4]} \Gamma_{a[4]}\right){}_\beta^{\;\alpha} \; , \qquad
\\ \label{RffL=}   R_{\alpha\beta}{}^{ab}=
 \left( -{2\over 3}
F^{ab[2]}\Gamma_{[2]} + {2i\over 3^. 5!} (\ast
F)^{ab[5]} \Gamma_{[5]} \right){}_{\alpha\beta}   \; . \qquad
\end{eqnarray}
Fixing the Wess--Zumino (WZ) gauge 
\begin{eqnarray}\label{WZgauge=} i_\Theta E^\alpha &:=&  \Theta^{\check{\beta}} E_{\check{\beta}}^\alpha (Z) = \Theta^\alpha\; , \quad  \nonumber \\ i_\Theta E^a&:=& \Theta^{\check{\alpha}} E_{\check{\alpha}}^{a}(Z)= 0\; , \quad  \nonumber \\
i_\Theta \omega^{ab}&:=&  \Theta^{\check{\alpha}} \omega_{\check{\alpha}}^{ab}=0\; , \qquad
\end{eqnarray}  one can solve the above constraints and find the complete expressions for the supervielbein and spin connection (see \cite{Claus} and refs therein)
\begin{widetext}
\begin{eqnarray}\label{Eua(th)=}
E^{\underline{a}}(x, \Theta) &=& e^{\underline{a}}(x) - 2i {\cal
D}\!\!\!^{^{^0}} \Theta^{\beta}  \; \sum\limits_{n=0}^{15}{1\over
(2n+2)!} \left((\Theta\Theta {\cal M})^n
\Gamma^{\underline{a}}\Theta\right)_\beta \; , \quad \\
\label{Ef(th)=}
 E^{\alpha}(x, \Theta) &=& {\cal D}\!\!\!^{^{^0}} \Theta^\alpha  + {\cal D}\!\!\!^{^{^0}} \Theta^{\beta}  \;
 \sum\limits_{n=1}^{16}{1\over  (2n+1)!}((\Theta\Theta {\cal M})^n )_\beta{}^\alpha  \; , \quad
\\  \label{omab(th)=}
 \omega^{\underline{a}\underline{b}}(x, \Theta) &=& \omega^{\underline{a}\underline{b}}(x) + {\cal D}\!\!\!^{^{^0}} \Theta^{\beta}  \;
 \sum\limits_{n=0}^{15}{1\over  (2n+1)!} \left((\Theta\Theta {\cal M})^n \right)_\beta{}^\gamma
R_{\gamma\alpha}{}^{\underline{a}\underline{b}}\,  \Theta^\alpha
 \; , \qquad
\end{eqnarray}
with
\begin{eqnarray}\label{calM:=}
(\Theta \Theta {\cal M})_\alpha{}^\beta :=  2i\,
(\Theta \Gamma^{{a}})_{\alpha}\; \Theta^{\gamma}T_{a\,
\gamma}{}^\beta - {1\over 4 } \Theta^{\gamma}R_{{\gamma}\alpha}{}^{ab}
\; (\Theta\Gamma_{{
a}{b}})^\beta
 \; . \qquad
\end{eqnarray}
Here $e^{\underline{a}}(x)= E^{\underline{a}}(x, 0)$ are the bosonic vielbein forms (\ref{ppw=e++}) and $\omega^{\underline{a}\underline{b}}(x)=
\omega^{\underline{a}\underline{b}}(x , 0)=dx^{{\mu}}
\omega_{{\mu}}^{\underline{a}\underline{b}}(x) $
is the purely bosonic limit of the $SO(1,9)$ ('spin')  connection, the nonvanishing components of which are given by Eq. (\ref{ppw=om--i}).
These are used to define ${\cal D}\!\!\!^{^{^0}} \Theta^\alpha$,
\begin{eqnarray}\label{cDth:=}
{\cal D}\!\!\!^{^{^0}} \Theta^\beta:= {D}\!\!\!^{^{^0}} \Theta^\beta - e^{\underline{a}}(x)\,  \Theta^\alpha  T_{\underline{\,a}\,\alpha}{}^\beta  =d\Theta^\beta -  \Theta^\alpha  {1\over 4} {\omega}{}^{\underline{a}\underline{b}}(x)   \Gamma_{\underline{a}\underline{b}\; \alpha}{}^{\beta} - e^{\underline{a}}(x)\,  \Theta^\alpha  T_{\underline{\,a}\,\alpha}{}^\beta \; . \qquad
\end{eqnarray}
Notice that $(\Theta \Theta {\cal M})_\alpha{}^\beta $ in (\ref{calM:=}) obeys $\Theta ^\alpha (\Theta \Theta {\cal M})_\alpha{}^\beta =0$ so that the WZ gauge conditions (\ref{WZgauge=}) are satisfied.

We will use the above expressions to study the pp-wave superspace, but notice that they can be also used to determine the supervielbein and spin connection forms of $AdS_4\times S^7$ and $AdS_4\times S^7$ superspaces. To be more specific, we  substitute (\ref{pp-w=F}) and (\ref{pp-w=F7}) and find that, for pp-wave superspace $\Sigma^{(11|32)}_{pp-w}$,
\begin{eqnarray}\label{Tbff=pp}
 E^a T_{{a} \, \beta }{}^\alpha & = &
 {i \mu \over 6} E^{++}
\left((I+{1\over 4}\Gamma^{--}\Gamma^{++})\Gamma^{123}\right){}_\beta^{\;\alpha} + {i \mu \over 6} E^{I}
\left(\Gamma^{++}\Gamma^{I}\Gamma^{1
23}\right){}_\beta^{\;\alpha} +
{i \mu \over 12} E^{\tilde{J}}
\left(\Gamma^{++}\Gamma^{\tilde{J}}\Gamma^{123}\right){}_\beta^{\;\alpha}
\; , \qquad
\\ \label{RffL=pp}
  R_{\alpha\beta}{}^{ab} \Gamma_{ab}{}_\gamma{}^\delta &= &
 -{4\mu \over 3} (\Gamma^{I}\Gamma^{123}){}_{\alpha\beta} (\Gamma^{++}\Gamma^{I}){}_\gamma{}^\delta
 - {i\mu \over 3\cdot 5!} \epsilon^{\tilde{I}\tilde{J}_1\ldots \tilde{J}_5} (\Gamma^{\tilde{J}_1\ldots \tilde{J}_5})
 {}_{\alpha\beta} (\Gamma^{++}\Gamma^{\tilde{J}}){}_\gamma{}^\delta
 -  \qquad \nonumber \\  &&  -
  {4\mu \over 3}
 (\Gamma^{++}\Gamma^{I})_{\alpha\beta} \epsilon_{IJK}\Gamma^{JK}{}{}_\gamma{}^\delta -
 {i\mu \over 6} \; {1\over 4!} \epsilon_{\tilde{I}\tilde{J}\tilde{J}_1\ldots \tilde{J}_4}((\Gamma^{++}\Gamma^{\tilde{J}_1\ldots \tilde{J}_4})
 {}_{\alpha\beta} (\Gamma^{\tilde{I}\tilde{J}}){}_\gamma{}^\delta
 \; . \qquad
\end{eqnarray}
\end{widetext} 
These expressions shall be used to specify (\ref{calM:=}). However, this gives a quite complex expression which does not result in essential simplification of the series in Eqs. (\ref{Ef(th)=})--(\ref{omab(th)=}).

The hope now is that, taking into account for the equations defining the center of energy motion of the mM$0$ system, or using a particular solution of these equations, one can simplify the expressions for pull--backs of the forms (\ref{Eua(th)=}), (\ref{Ef(th)=}), (\ref{omab(th)=}), (\ref{calM:=}), (\ref{Tbff=pp}), (\ref{RffL=pp}) to ${\cal W}^{(1|16)}$  in such a way that the equations for the relative motion of mM$0$ constituents,  
(\ref{M0:DtPsi=}), (\ref{M0:XiDXi=}) and (\ref{M0:DDXi=}) become manageable.

\section{Multiple M$0$-branes in supersymmetric  pp-wave background and BMN model equations }
\setcounter{equation}0

In \cite{mM0=PLB,mM0=PRL,mM0=PRD} the center of energy motion of the mM$0$ system are described by the same equations as the motion of a single M$0$-brane. More precisely, the embedding of the center of energy  superspace ${\cal W}^{(1|16)}$ into the target 11D supergravity superspace is assumed to be described by the same superembedding equations as the embedding of  the worldline superspace of a single M$0$.  Consequently, to study the mM$0$ system in the 11D pp-wave background it is necessary to find a(t least a particular) solution of the superembedding equation describing  embedding of ${\cal W}^{(1|16)}$  into the 11D pp-wave superspace. 

\subsection{Worldline superspace ${\cal W}^{(1|16)}$ and superfield equations of M$0$-brane}

In the superembedding description the coordinate functions of M$0$--brane (or of the center of energy of the 
mM$0$ system) are
superfields  $\hat{Z}^{\underline{M}}(\tau, \eta^{q})=(\hat{X}{}^{ \underline{\mu}}(\zeta)\, , \hat{\theta}^{\check{ \underline{\alpha}}}(\tau, \eta^{q}))$ which determine the embedding of the (center of energy) {\it worldline superspace} ${\cal W }^{(1|16)}$, Eq. (\ref{M0:W1-16=teta}), into the target 11D superspace $\Sigma^{(11|32)}$,  
\begin{eqnarray}
\label{WinS} & {\cal W }^{(1|16)}\in \Sigma^{(11|32)} : \quad
Z^{ {M}}=  \hat{Z}^{ {M}}(\tau ,  \eta^{\check{q}})\;  . \qquad 
\end{eqnarray}
Using the gauge symmetries one can also specify the form of the coordinate superfields 
\begin{eqnarray}
\label{Z=X+Th} & 
  \hat{Z}^{ {M}}(\tau ,  \eta^{\check{q}})=  (\hat{X}^{ {\mu}}(\tau ,  \eta), \hat{\Theta}{}^{\check\alpha}(\tau, \eta))
 \;  . \qquad
\end{eqnarray} 
In particular, the fermionic coordinate function can be presented in the form
\begin{eqnarray}\label{Th=eta+th}
\hat{\Theta}{}^{\alpha}(\tau, \eta) = \eta^{+q}v_q^{-\alpha}+ \Theta^{
-q}(\tau, \eta) v_q^{+\alpha}\; ,
\qquad
\end{eqnarray}
which separates explicitly the 16 component (SO(9) spinor)  Goldstone fermion superfield $\Theta^{
-q}(\tau, \eta)= \theta^{
-q}(\tau)+ \propto \eta^{+p}= \hat{\Theta}{}^{\alpha}(\tau, \eta) v_{\alpha }{}^-_q$ (associated with the center of energy motion) and identify the remaining 16 components of the 32 component  $\hat{\Theta}{}^{\alpha}(\tau, \eta)$ with Grassmann coordinates of the worldvolume superspace, $ \hat{\Theta}{}^{\alpha}(\tau, \eta) v_{\alpha }{}^+_q=\eta^{+q}\;$  \footnote{Notice that in the Wess--Zumino gauge (\ref{WZgauge=}) the index of the fermionic coordinate of superspace is identified with the $Spin(1,10)$ index, which makes possible to write Eq. (\ref{Th=eta+th}) in a generic curved supergravity superspace. Similarly, after fixing the gauge (\ref{Th=eta+th}), the worldline superspace fermionic coordinate $\eta^{+p}$ is transformed nontrivially under the $SO(1,1)$ transformations acting on spinor moving frame superfields $v_q^{\pm\alpha}$.}. Actually, it is often convenient to have a bit more complicated gauge fixing $ \hat{\Theta}{}^{\alpha}(\tau, \eta) v_{\alpha }{}^+_q=\eta^{+p}A_{pq}\;$ with some invertible  matrix $A_{pq}=A_{pq}(\hat{x}, \hat{\Theta}^-)$ (see Eq. (\ref{Th+q=f+qx}) below). 

The coordinate superfields (\ref{Z=X+Th}) are required to obey the superembedding equation which can be written with the use of moving frame superfield $u^{= a}(\tau,\eta)$, obeying (\ref{u--u--=0}), as
\begin{eqnarray}
\label{Eua=e++u--}
  \hat{E}{}^{{a}}:= d\hat{Z}^M(\tau,\eta) E_M{}^{{a}}(\hat{Z}(\tau,\eta)) = {1\over 2}{\cal E}^{\#}u^{= a}(\tau,\eta)\; .
\qquad
\end{eqnarray}
Here $\hat{E}{}^{{a}}:= {E}{}^{{a}}(\hat{Z})$ is the pull--back of the bosonic vielbein of target superspace to ${\cal W }^{(1|16)}$ and
${\cal E}^{\#}=d\tau {\cal E}^{\#}_\tau (\tau, \eta)+ d\eta^{\tilde{q}} {\cal E}^{\#}_{\tilde{q}} (\tau, \eta)$ is the bosonic supervielbein of ${\cal W }^{(1|16)}$. This is induced by embedding, {\it i.e.} it is expressed through $\hat{E}{}^{{a}}$ and the moving frame superfield by the equation 
\begin{eqnarray}
\label{Ea=e++u}
{\cal E}^{\#}= \hat{E}{}^{{a}}u_a^{\#}\; \qquad
\end{eqnarray}
which follows from (\ref{Eua=e++u--}) and (\ref{u++u++=0}). The fermionic supervielbein form of ${\cal W }^{(1|16)}$ is denoted by ${\cal E}^{+q}=d\tau {\cal E}^{+q}_\tau (\tau, \eta)+ d\eta^{\tilde{p}} {\cal E}^{+q}_{\tilde{p}} (\tau, \eta)$ and is expressed through the pull--back of the fermionic supervielbein  to ${\cal W }^{(1|16)}$, $\hat{E}{}^{{\alpha}}:= {E}{}^{{\alpha}}(\hat{Z})$, and spinor moving frame superfields by one of the projections of the equation 
\begin{eqnarray}
\label{Euf=e+qv-q}
  \hat{E}{}^{{\alpha}}:= d\hat{Z}^M(\tau,\eta) E_M{}^{{\alpha}}(\hat{Z}(\tau,\eta)) = {\cal E}^{+q} v_q^{-\alpha}(\tau,\eta)\; ,
\qquad
\end{eqnarray}
namely 
\begin{eqnarray}\label{hEfv+=e+q}
{\cal E}^{+q} = \hat{E}{}^{{\alpha}}v_{\alpha}{}^+_q\; .
\qquad
\end{eqnarray}
The other projection,
\begin{eqnarray}
\label{hEfv-=eqm}
  \hat{E}{}^{{\alpha}} v_{\alpha}{}^-_q(\tau,\eta) = 0\; ,
\qquad
\end{eqnarray}
gives the superfield generalization of the fermionic equations of motion of the M$0$-brane.

Actually, the M$0$--brane equations of motion can be expressed by stating the covariant constancy of the light--like vector $u^{a=}$, 
\begin{eqnarray}
\label{Dua--=0}
Du^{a=}(\tau, \eta):= du^{a=} - u_b^{=}\hat{\omega}{}^{ba}- 2u^{a=} \omega^{(0)}(\tau, \eta)= 0\; . \quad
\end{eqnarray}
This supports the above statement on that the M$0$--brane  dynamics can be completely described in terms of the moving frame superfields.

Following \cite{mM0=PLB,mM0=PRL,mM0=PRD}, we describe the center of energy motion of the mM$0$ system by the equations which coincide with the equations of motion of a single M$0$-brane. This implies that the above equations are used to describe the center of energy superspace ${\cal W}^{(1|16)}$ of mM$0$ system.

As far as our main interest here is in relation with the BMN matrix model, for simplicity we will not try to find general solution of the equations of the center of energy motion of the mM$0$ system, but rather chose a particular solution of the equations of the superembedding approach, specifying the embedding of a particular worldline superspace ${\cal W}_0^{(1|16)}$ in the pp-wave superspace $\Sigma^{(11|32)}_{pp-w}$ described in the previous section. 

\subsection{Worldline superspace ${\cal W}_0^{(1|16)}$ describing a particular 1/2 BPS state of M$0$-brane in  $\Sigma^{(11|32)}_{pp-w}$}

Our  ${\cal W}_0^{(1|16)}$ superspace  is characterized by vanishing of the Goldstone fermion superfield in (\ref{Th=eta+th}), so that
\begin{eqnarray}\label{Th=Th+q}
\hat{\Theta}{}^{\alpha}(\tau, \eta) = \Theta^{+q}(\tau, \eta)v_q^{-\alpha}\; ,
\qquad
\end{eqnarray}
by vanishing  of nine bosonic Goldstone superfields
\begin{eqnarray}\label{hxI=0}
\hat{x}{}^{I}(\tau, \eta) = 0\; ,  \qquad \hat{x}{}^{\tilde{J}}(\tau, \eta) = 0\; ,  \qquad
\qquad
\end{eqnarray}
and by constant moving frame vectors
\begin{eqnarray}\label{u--=const}
u_a^{=}&=& \delta_a^{--}\equiv  \delta_a^{0}-\delta_a^{10} \; , \qquad
u_a^{\#}= \delta_a^{++}\equiv  \delta_a^{0}+\delta_a^{10}\; , \qquad \nonumber \\
&& u_a^{i}= \delta_a^{i} \; . \qquad
\end{eqnarray}
This ${\cal W}_0^{(1|16)}$ describes a particular supersymmetric ground state of a center of energy of the mM$0$-system (or of the single M$0$--brane) in the 11D pp-wave  background. 

Eq. (\ref{u--=const}) implies that the spinor moving frame superfields are also constant,
\begin{eqnarray}\label{v-q=const}
v_q^{-\alpha}=\delta_{+q}^{\alpha}\; \qquad v_q^{+\alpha}=\delta_{-q}^{\alpha}
\; , \qquad
\end{eqnarray}
obeying
\begin{eqnarray}\label{v+G++=00}
v_q^{+\alpha}\Gamma^{++}_{\alpha\beta}=0 \; , \qquad  v_q^{+\alpha }\Gamma^{--}_{\alpha\beta}= 2v_{\beta q}^{\; -}\; , \qquad  \nonumber \\
v_q^{+\alpha }\Gamma^{i}_{\alpha\beta}= \gamma^i_{qp}v_{\beta p}^{\; +}\; , \qquad \\ \label{v-G--=0}
v_q^{-\alpha }\Gamma^{++}_{\alpha\beta}= 2v_{\beta q}^{\; +} \; , \qquad v_q^{-\alpha }\Gamma^{--}_{\alpha\beta}=0 \; , \qquad \nonumber \\
v_q^{-\alpha }\Gamma^{i}_{\alpha\beta}= \gamma^i_{qp}v_{\beta p}^{\; -}  v_{\beta q}^{\; +} \;
 . \qquad
\end{eqnarray}
Notice that the (constant) frame (\ref{u--=const}) is oriented in such a way that
\begin{eqnarray}\label{hF++=vacuum} && \hat{F}_{\# ijk}= -6\delta^{[i}_1 \delta^{j}_2 \delta^{k]}_3F_{++ 123}=
-{3\mu } \delta^{[i}_1 \delta^{j}_2 \delta^{k]}_3 \, ,  \qquad  \\ \label{hRIJ=vacuum} && \hat{R}_{\# I J \# }=4\left({\mu\over 3}\right)^2 \delta^{IJ}\; , \quad
 \hat{R}_{\# \tilde{I}\tilde{J} \#}= 4\left({\mu\over 6}\right)^2 \delta^{\tilde{I}\tilde{J}} . \qquad
\end{eqnarray}  Using these equations one can easily check  that the BPS conditions (\ref{DpTq=0->R}) and
(\ref{DpTq=0->FF3}) are obeyed. Below we will also see that Eq. (\ref{DpTq=0->DF}) is satisfied so that our particular solution of the superembedding approach equations preserves 1/2 of the target space supersymmetry.

The coordinate function $\hat{x}^{++}$ can be identified with the (generalization of the) particle proper time,
\begin{eqnarray}\label{hx++=tau}
\hat{x}^{++}(\tau, \eta)=\tau , \qquad
\end{eqnarray}
while the remaining bosonic Goldstone superfield $\hat{x}^{--}(\tau, \eta)$ can take an arbitrary constant value, $\partial_\tau \hat{x}^{--}(\tau, \eta)=0$. The nonvanishing components $\Theta^{+q}$ of the Grassmann superfield $\Theta^\alpha$ are related to the Grassmann coordinates $\eta^{+q}$ of the worldvolume superspace ${\cal W}^{(1|16)}$ by
\begin{eqnarray}\label{Th+q=f+qx}
\hat{\Theta}{}^{+q}(\tau ,\eta)= \eta^{+p}\; \left(exp\left\{-{\mu\over 6} \, \hat{x}^{++}\gamma^{123}\right\}\right)_{pq}\; . \qquad
\end{eqnarray}

One can check that this configuration solves the equations of the superembedding approach and, hence, describes a particular solution of the equations of motion of single M$0$-brane in $\Sigma^{(11|32)}_{pp-w}$. Indeed, it is characterized by
\begin{eqnarray}\label{eIvac=0} e^I(\hat{x})=0\, , \quad e^I(\hat{x})=0\; , \quad e^{++}(\hat{x})=d\hat{x}^{++}\; ,  \qquad \\ \nonumber
\omega^{ab}(\hat{x})=0\; , \qquad
\end{eqnarray}
so that, at zero order in Grassmann coordinate $\eta^{+q}$, the equations of motion (\ref{Dua--=0}) are satisfied.

Then, Eqs. (\ref{eIvac=0}) and (\ref{Tbff=pp}) result in 
\begin{eqnarray}\label{eaTaff=}
\hat{e}^a T_{a\beta}{}^\alpha = -{\mu\over 6}d\hat{x}^{++}\; (\Theta^{+}\gamma^{123})_q v^{-\alpha}_q \; , \qquad
\end{eqnarray}
so that calculating the pull--back of ${\cal D}^{\!\!\! ^0}\Theta^\alpha$ in (\ref{cDth:=}), one finds
\begin{eqnarray}\label{cDhTh=dfv-}
{\cal D}^{\!\!\! ^0}\hat{\Theta}{}^\alpha= d\eta^{+p}\; \left(exp\left\{-{\mu\over 6} \, \hat{x}^{++}\gamma^{123}\right\}\right)_{pq}v^{-\alpha}_q\; , \qquad
\end{eqnarray}
which implies
\begin{eqnarray}\label{cDhThv-=0}
{\cal D}^{\!\!\! ^0}\hat{\Theta}{}^\alpha v_{\alpha q}{}^-=0\; , \qquad
\\ \label{cDhThv+=}
{\cal D}^{\!\!\! ^0}\hat{\Theta}{}^\alpha v_{\alpha q}{}^+= d\eta^{+p}\; \left(exp\left\{-{\mu\over 6} \, \hat{x}^{++}\gamma^{123}\right\}\right)_{pq}\; . \qquad
\end{eqnarray}

At this stage it is important to notice that the pull--back to ${\cal W}_0^{(1|16)}$ of $\Theta\Theta {\cal M}$, defined in (\ref{calM:=}), (\ref{Tbff=pp}) and (\ref{RffL=pp}), has the following block-diagonal structure
\begin{eqnarray}\label{hThhThcM=0}
(\hat{\Theta}\hat{\Theta} {\cal M})_\beta{}^\alpha= v_{\beta p}{}^+ (S^T(\eta\eta{\cal M})^{-+}S)_{pq}v_q^{-\alpha} + \qquad \nonumber \\ + v_{\beta p}{}^- (S^T(\eta\eta{\cal M})^{+-}S)_{pq}v_q^{+\alpha}  \; .  \qquad
\end{eqnarray}
The explicit form of the matrices $S$, $(\eta\eta{\cal M})^{-+})_{pq}$ and $(\eta\eta{\cal M})^{+-})_{pq}$ (which can be found in the Appendix, Eqs. (\ref{S=exp}), (\ref{ffcM-+=}) and (\ref{ffcM+-=}))  is not needed to check that
Eqs. (\ref{hThhThcM=0}) and (\ref{Ef(th)=}) imply that $E^{\alpha}v_{\alpha q}{}^-\propto
{\cal D}^{\!\!\! ^0}\hat{\Theta}{}^\alpha v_{\alpha p}{}^- \, $. Then the (superfield generalization of the) fermionic equations of motion, Eq. (\ref{hEfv-=eqm}) are obeyed as a consequence of (\ref{cDhThv-=0}),
\begin{eqnarray}\label{Efv-q=0}
 \hat{E}^{\alpha} v_{\alpha q}^{\; -} = {\cal D}\!\!\!^{^{^0}} \Theta^\alpha v_{\alpha p}^{\; -} \left(\delta_{pq} +
 \sum\limits_{n=1}^{8}{((\eta\eta{\cal M})^{+-})^n )_{pq}\over  (2n+1)!}\right)=0  \, . \nonumber \\ {}
\end{eqnarray} 
Furthermore, using (\ref{cDhThv-=0}) one finds that the superembedding equations is satisfied to all orders in $\eta^{+q}$, this is to say that $\hat{E}^{I}=0$, $\hat{E}^{\tilde{J}}=0$ and  $\hat{E}^{--}=0$.

The fermionic  and bosonic supervielbein forms of  the worldvolume superspace read
\begin{eqnarray}\label{cE++=Pi+}
&& {\cal E}^{\#}= d\hat{x}^{++} - 2i d\eta^{+q}\, \eta^{+q} - 4i d\eta^{+p}
(\eta\eta{\cal K})_{pq}\eta^{+q} , \qquad
 \\ \label{e+q=Efv+q}
&& {\cal E}^{+q}=d\eta^{+p} {\Xi}^{pq}(\tau,\eta)    \;   \quad
\end{eqnarray}
(see  Eqs. (\ref{Kpq:=})  and (\ref{calSpq:=}) in the Appendix  for the explicit form of  $(\eta\eta{\cal K})_{pq}$ and $ {\Xi}^{pq}(\tau,\eta)$).
This implies that the covariant derivatives for scalar superfields are
\begin{eqnarray}\label{nabla++=}
&& \nabla_{\#}= \hat{\partial}_{++}  \; , \qquad  \nonumber \\ && \nabla_{+q}= (L^{-1})^{qp}(D_{+p} +4i (\eta\eta{\cal K})_{pp^\prime}\eta^{+p^\prime} \hat{\partial}_{++} ) \; , \qquad
\end{eqnarray}
where
\begin{eqnarray}\label{D+q=} D_{+q}=  \partial_{+q} +2i \eta^{+q} \hat{\partial}_{++}\;  , \qquad  \partial_{+q} := {\partial \over \partial \eta^{+q}}\; , \qquad \\ \nonumber \hat{\partial}_{++}  := {\partial\over \partial \hat{x}^{++}}=
{1\over {\partial_\tau \hat{x}^{++}} }{\partial \over \partial \tau} \;  \qquad
\end{eqnarray}
are derivatives covariant  with respect to the flat superspace supersymmetry.

The pull--back of the spin connection forms to ${\cal W}^{(1|16)}_{0}$ is characterized  by 
\begin{eqnarray}\label{hom--i=0}
\omega^{--i}(\hat{x},\hat{\Theta})=0\; , && \quad \omega^{++i}(\hat{x},\hat{\Theta})=0\; , \quad \nonumber \\ \omega^{--\; ++}(\hat{x},\hat{\Theta})=0\; , && \quad  \omega^{{I}\tilde{J}}(\hat{x},\hat{\Theta})= 0 \; , \quad
\\ \label{homIJ=0} \omega^{IJ}(\hat{x},\hat{\Theta})&=&d\eta^{+q}\omega_{+q}^{IJ}(\hat{x},\hat{\Theta}) \; , \quad \\
\label{homtItJ=0} \omega^{\tilde{I}\tilde{J}}(\hat{x},\hat{\Theta})&=&d\eta^{+q}\omega_{+q}^{\tilde{I}\tilde{J}}(\hat{x},\hat{\Theta})  \qquad 
\end{eqnarray}
(see  Eqs. (\ref{homIJ=df}) and (\ref{homtItJ=df}) in Appendix for  explicit form of $\omega_{+q}^{IJ}(\hat{x},\hat{\Theta})$ and $\omega_{+q}^{\tilde{I}\tilde{J}}(\hat{x},\hat{\Theta})$) so that  
it is easy to check that the (superfield generalization of the) M$0$--brane bosonic equation, Eq. (\ref{Dua--=0}), is satisfied.

Furthermore, using the fact that the moving frame vectors characterizing our solution are constant, $du^{=}_a=0$, $du^{\#}_a=0$, $du^{i}_a=0$, together with Eqs. (\ref{Dui}) and   (\ref{hom--i=0})--(\ref{homtItJ=0}), we find
that $\Omega^{\# i}=0$ ($\Omega^{= i}=0$  is equivalent to (\ref{Dua--=0})) as well as the explicit form of the $SO(9)$ and $SO(1,1)$ connection. These are characterized by 
\begin{eqnarray}\label{AIJ=0}
\Omega^{(0)}=0\; , \qquad A^{I\tilde{J}}=0 \; , \qquad  \nonumber \\ A^{IJ}=
d\eta^{+q}  A_{+q}^{IJ} \; , \qquad 
A^{\tilde{I}\tilde{J}}=d\eta^{+q}  A_{+q}^{\tilde{I}\tilde{J}}\;  \qquad
\end{eqnarray}
which imply $\Omega_\#^{(0)}=0$ and $A_\#^{ij}=0$ so that
\begin{eqnarray}\label{D++=}
D_\#= \nabla_\# = \hat{\partial}_{++}\; . \qquad
\end{eqnarray}

Now we  see that the third BPS equation  (\ref{DpTq=0->DF}) is satisfied just because $F_{\# ijk}$ in Eq.  (\ref{hF++=vacuum}) is a constant. This completes the proof of the fact that  our particular solution of the superembedding approach equations preserves 1/2 of the  supersymmetry of the 11D pp-wave background. This 16 preserved supersymmetry of the target superspace are identified with the supersymmetry of the worldline superspace ${\cal W}_0^{(1|16)}$ described by this solution so that the equations of the relative motion of the mM$0$ system, defined on this superspace in a manifestly covariant way, are invariant under this 16 parametric supersymmetry by construction.

\subsection{Multiple M$0$-brane equations  in 11D pp-wave background and BMN matrix model equations}

Now we are ready to write explicitly the equations of the relative motion of the constituents of multiple M$0$--brane system the center of energy of which moves in the pp--wave superspace in the above described particular manner. These read
\begin{widetext}
\begin{eqnarray}\label{M0:DtPsi=pp}
 \hat{\partial}_{++}{\Psi}_{q}+   {\mu\over 4}  \gamma^{123}_{qp}\Psi_{p}&=&- {1\over 4} \gamma^i_{qp} \left[ {\bf X}^i\, , \,
\Psi_{p} \right]
\, ,
\\ \label{M0:XiDXi=pp}
 \left[ {\bf X}^i\, , \, \hat{\partial}_{++}{\bf X}^i\,  \right] &=& 4i \left\{\Psi_{q}\, , \, \Psi_{q} \right\}  \;  \qquad
\\ \label{M0:DDXI=}
\hat{\partial}_{++}\hat{\partial}_{++}{\bf X}^I + \left({\mu \over 3}\right)^2  {\bf X}^I &=&
{1\over 16} \left[ {\bf X}^j , \left[ {\bf X}^j ,  {\bf X}^I\,  \right]\right]+ i\gamma^I_{qp} \left\{\Psi_{q}\, , \, \Psi_{p} \right\} -
{\mu \over 8} \epsilon^{IJK} \left[ {\bf X}^J , \, {\bf X}^K\,  \right]   \, , \quad \\
\label{M0:DDXtJ=}
\hat{\partial}_{++}\hat{\partial}_{++}{\bf X}^{\tilde{J}} +   \left({\mu \over 6}\right)^2  {\bf X}^{\tilde{J}} &=&
{1\over 16} \left[ {\bf X}^i , \left[ {\bf X}^i ,  {\bf X}^{\tilde{J}}\,  \right]\right]+ i\gamma^{\tilde{J}}_{qp} \left\{\Psi_{q}\, , \, \Psi_{p} \right\}   \, .
\end{eqnarray}
\end{widetext}
These equations coincide with the ones which can be obtained by varying the BMN action \cite{BMN} up to the fact that they are formulated for the traceless matrices.

The trace parts of the matrices in  \cite{BMN} should be related with the center of energy motion of the mM$0$ system. In our approach that is described separately by the geometry of the embedding of the center of energy worldvolume superspace ${\cal W}^{(1|16)}$ into $\Sigma^{(11|32)}_{pp-w}$. Thus to find the equations of motion for the center of energy coordinate functions (which are essentially center of energy  Goldstone bosons and fermions), one should go beyond the ground state solution of the superembedding equation, which we have used above. This will be the subject of the forthcoming paper. Here we will restrict ourself by showing that the complete set of the the BMN equations is reproduced by the mM$0$ equations in the leading approximation on the center of energy Goldstone fields.

\subsection{BMN model as an approximation description of  mM$0$ system in 11D  pp-wave background }
\label{U1} 

To this end let us consider the leading components ($\eta^{+q}=0$ limits) of the superembedding equation (\ref{Eua=e++u--}) and of other equations of superembedding approach which follows from that, Eqs.  (\ref{hEfv-=eqm}) and (\ref{Dua--=0}). These can be solved  by the ansatz
\begin{eqnarray}\label{Th=th-v+0}
\hat{\Theta}{}^{\alpha}(\tau, \eta) =  \theta^{
-q}(\tau) v_q^{+\alpha}\;
\qquad
\end{eqnarray}
and
\begin{eqnarray}\label{u--=solution}
u_a^{=}&=& \delta_a^{--}+ {1\over 4} k^{--i} k^{--i}\delta_a^{++}- k^{--i}\delta_a^{i}\; , \qquad  \nonumber \\
u_a^{\#}&=& \delta_a^{++}\equiv  \delta_a^{0}+\delta_a^{10}\; , \qquad \nonumber \\
u_a^{i}&=& \delta_a^{i} -{1\over 2} k^{--i} \delta_a^{++}\; . \qquad
\end{eqnarray}
Substituting (\ref{u--=solution}) and (\ref{Th=th-v+0}) into Eqs.
(\ref{Eua=e++u--}), (\ref{hEfv-=eqm}) and (\ref{Dua--=0}), one finds that
$k^{--i}=2\hat{\partial}_{++}\hat{x}^i$ and that the bosonic and fermionic Goldstone fields
have to obey the following equations of motion
\begin{eqnarray}\label{ddx=xmu}
&& \hat{\partial}_{++}\hat{\partial}_{++} \hat{x}^{I}(\tau) + \left({\mu\over 3}\right)^2\, \hat{x}^I(\tau )=0 \; , \qquad \nonumber \\ && \hat{\partial}_{++}\hat{\partial}_{++} \hat{x}^{\tilde{J}}(\tau )+\left({\mu\over 6}\right)^2\, \hat{x}{}^{\tilde{J}}(\tau )=0\; ,  \qquad
\\ \label{d++th-q=}
&& \hat{\partial}_{++}\theta^{-q}(\tau) - {\mu\over 6}\theta^{-p}(\tau)\gamma^{123}_{pq} =0 \; .
\end{eqnarray}

Now, let us turn back to the relative motion equations (\ref{M0:DtPsi=pp}), (\ref{M0:XiDXi=pp}), (\ref{M0:DDXI=}), (\ref{M0:DDXtJ=}) which are written for traceless matrices. The BMN action produces the same equations but for tracefull matrices.
As trace parts of the matrices do not contribute to the commutator terms, the traceless parts of BMN equations coincide with (\ref{M0:DtPsi=pp}), (\ref{M0:XiDXi=pp}), (\ref{M0:DDXI=}), (\ref{M0:DDXtJ=}).

The trace part of the BMN counterparts of equations  (\ref{M0:DDXI=}) and  (\ref{M0:DDXtJ=}) coincides with Eqs. (\ref{ddx=xmu}). The trace part of the BMN version of the fermionic equations (\ref{M0:DtPsi=pp}) seems to be different from (\ref{d++th-q=}): it
describes a relativistic fermion of mass $\mu\over 4$ rather then $\mu\over 6$ in (\ref{d++th-q=}). However, in $d=1$ field theory the fermionic mass is a matter of convenience in the sense that the value of mass can be changed by an appropriate (coordinate dependent) field redefinition.
Indeed the field $\psi^{-q}:= \theta^{-p} exp\left\{\left({\mu\over 12}\hat{x}^{++}\gamma^{123}\right)\right\}_{pq}$ obeys the equation for massive fermion of mass $\mu/4$,
\begin{eqnarray}\label{d++psi-q=}
\hat{\partial}_{++}\psi^{-q}(\tau) = {\mu\over 4}\psi^{-p}(\tau)\gamma^{123}_{pq} \;
\end{eqnarray}
which coincides with the trace part of the BMN version of Eq. (\ref{M0:DtPsi=pp}).

This allows us to conclude that the mM$0$ equations in general 11D supergravity background obtained in \cite{mM0=PRL,mM0=PRD}, when specified for the case of completely supersymmetric pp-wave solution, does reproduce the equations of BMN model as an approximation. This is a leading approximation in the decomposition on powers of the center of energy Goldstone fields (bosonic $\hat{x}^I$,  $\hat{x}^{\tilde{J}}$ and fermionic $\theta^{-q}$) which is made over the supersymmetric vacuum solution of the equations of center of energy motion described by Eqs. (\ref{hxI=0}), (\ref{u--=const}), (\ref{hx++=tau}), (\ref{Th=Th+q}), (\ref{Th+q=f+qx}).

\section{Conclusion}

In this paper we have begun the program of the studying and developing applications of the Matrix model equations in general 11D supergravity background \cite{mM0=PRL,mM0=PRD} by specifying them for particular 11D background which are interesting in M-theoretical perspective. We have begun by the case of completely supersymmetric 11D pp-wave background, which is natural as far as a Matrix model in this background have been proposed nine years ago by Berenstein, Maldacena and Nastase \cite{BMN}. So the good check of the multiple M$0$-brane equations of \cite{mM0=PRL,mM0=PRD} as equations for matrix model in an arbitrary 11D supergravity background is to check whether they can reproduce the equations of BMN model  \cite{BMN}.

We have shown that, when the Goldstone fields of the center of energy motion of multiple M$0$-brane (mM$0$) system are set to zero, and the center of energy motion preserves 1/2 of the supersymmetries of the pp-wave background,
the equations describing relative motion of the mM$0$ constituents coincide with the BMN equations, but for written for traceless matrix fields. Furthermore, we have shown that the complete mM$0$ equations in pp-wave superspace actually reproduce the BMN equations in the leading approximation on Goldstone bosons and fermions describing the center of energy motion (when these are defined as excitations over the above mentioned 1/2 BPS vacuum solution).

The complete accounting of the contribution of the center of energy Goldstone fields $\hat{x}^I$,  $\hat{x}^{\tilde{J}}$ and $\theta^{-q}(\tau)$ into the equations of relative motion of mM$0$ constituent requires  a complete description of an arbitrary center of energy superspace ${\cal W}^{(1|16)}$, this is to say it requires a more general solution of the superembedding approach equations in the case of pp-wave target superspace $\Sigma^{(11|32)}_{pp-w}$. Then the equation of relative motion could get modified by contributions of these center of energy Goldstone fields and thus deviate from the BMN equations. We hope to turn to this issue in the future publication.

\bigskip
\acknowledgements{
The author is thankful to Dima Sorokin for useful discussions, to Djordje Minic for the interest to this work and useful conversations, and to the organizers of the Miami 2011 Conference, especially to Thomas Curtright, Jo Ann Curtright and Luca Mezincescu, for the hospitality at Fort Lauderdale on a final stage of this work which was supported in part by the research grants FIS2008-1980 from the MICINN of Spain, the Basque Government Research Group Grant ITT559-10 and by the UPV/EHU under program UFI 11/55.}

\section*{Appendix. }
\renewcommand{\theequation}{A.\arabic{equation}}

In this Appendix we collect some technical details.

The explicit form of the matrices entering  Eq. (\ref{hThhThcM=0}) is given by
\begin{widetext}
\begin{eqnarray}\label{S=exp}
S_{pq}&:=&\left(exp\left\{-{\mu\over 6} \, \hat{x}^{++}\gamma^{123}\right\}\right)_{pq}
\; , \qquad  \\ \label{ffcM-+=}
(\eta\eta{\cal M})^{-+}_{pq} &=& -{2i\mu\over 3} \eta^{+p}  ({\eta}^+\gamma^{123})_q
-  {i\mu\over 3} \epsilon^{IJK} (\eta^{+}\gamma^{I})_p
  (\eta^{+}\gamma^{JK})_{q}
  -  {i\mu\over 12} \,  {1\over  4!} \epsilon_{\tilde{I}\tilde{J}\tilde{J}_1\ldots \tilde{J}_4}(\eta^{+}\gamma^{\tilde{J}_1\ldots \tilde{J}_4})_{p}({\eta}^+\gamma^{\tilde{I}\tilde{J}})_q \; ,
 \\ \label{ffcM+-=}
(\eta\eta{\cal M})^{+-}_{pq} &=& -
{2i\mu\over 3} (\eta^{+}\gamma^{I})_p  ({\eta}^+\gamma^{I}\gamma^{123})_q-
{2i\mu\over 3} (\eta^{+}\gamma^{I}\gamma^{123})_p  ({\eta}^+\gamma^{I})_q
-  {i\mu\over 3} (\eta^{+}\gamma^{\tilde{J}})_p  ({\eta}^+\gamma^{\tilde{J}}\gamma^{123})_q - \qquad \nonumber \\
 && -  {i\mu\over 6} \,  {1\over  5!} \epsilon_{\tilde{J}\tilde{J}_1\ldots \tilde{J}_5}(\eta^{+}\gamma^{\tilde{J}_1\ldots \tilde{J}_5})_{p}({\eta}^+\gamma^{\tilde{J}})_q
\end{eqnarray}

The matrices entering Eqs. (\ref{cE++=Pi+}) and (\ref{e+q=Efv+q}) read
\begin{eqnarray}\label{Kpq:=}
(\eta\eta{\cal K})_{pq}&:=&
 \sum\limits_{n=1}^{7}{1\over  (2n+2)!}((\eta\eta{\cal M})^{-+})^n )_{pq}  \; , \quad
\\ \label{calSpq:=}
{\Xi}^{pq}(\tau,\eta) &=:&\left(\delta_{pp^\prime} +
 \sum\limits_{n=1}^{8}{1\over  (2n+1)!}((\eta\eta{\cal M})^{-+})^n )_{pp^\prime}\right)\left(exp\left\{-{\mu\over 6} \, \hat{x}^{++}\gamma^{123}\right\}\right)_{p^\prime q}    \;  . \quad
\end{eqnarray}

The nonvanishing components of the pull--back of pp-wave superspace spin connection to ${\cal W}_0^{(1|6)}$   
(see  Eqs. (\ref{homIJ=0}) and (\ref{homtItJ=0})) are  
\begin{eqnarray}\label{homIJ=df}
\omega^{IJ}(\hat{x},\hat{\Theta})={8i\mu\over 3 }\epsilon^{IJK} d\eta^{+q}
 \sum\limits_{n=1}^{7}{1\over  (2n+2)!}((\eta\eta{\cal M})^{-+})^n\gamma^K\eta^+)_q\; , \qquad \\  \label{homtItJ=df}
\omega^{\tilde{I}\tilde{J}}(\hat{x},\hat{\Theta})={\mu\over 6 }\; {1\over 4!} \epsilon_{\tilde{I}\tilde{J}\tilde{J}_1\ldots \tilde{J}_4} d\eta^{+q}
 \sum\limits_{n=1}^{7}{1\over  (2n+2)!}((\eta\eta{\cal M})^{-+})^n \gamma^{\tilde{J}_1\ldots \tilde{J}_4}\eta^+)_q\;  . \quad
\end{eqnarray}
The   $SO(1,1)$ connection on ${\cal W}_0^{(1|6)}$ and of the connection on the normal bundle over ${\cal W}_0^{(1|6)}$ read 
\begin{eqnarray}\label{AIJ=}
&& \Omega^{(0)}=0\; , \qquad \Omega^{\# I}=0=\Omega^{\# \tilde{J}}=0 \; ,  \qquad \nonumber \\   A^{I\tilde{J}}=0 \; , && \quad A^{IJ}={8i\mu\over 3}\epsilon^{IJK}
d\eta^{+q} {\Xi}^{qp}(\gamma^{K} \eta^{+})_p  \; , \qquad  
A^{\tilde{I}\tilde{J}}= {\mu\over  6} \; {1\over 4!}  \epsilon^{\tilde{I}\tilde{J}\tilde{J}_1\ldots \tilde{J}_4}
d\eta^{+q} {\Xi}^{qp}(\gamma^{\tilde{J}_1\ldots \tilde{J}_4}\eta^{+})_{p}\; , \qquad
\end{eqnarray}
where ${\Xi}^{qp}$ is the invertible matrix presented in  Eq. (\ref{calSpq:=}). 
\end{widetext}

\end{document}